\documentclass{iopart}
\usepackage{amssymb}
\usepackage{siunitx}  
\usepackage{graphicx}
\usepackage{hyperref}
\usepackage[utf8]{inputenc}
\usepackage[T1]{fontenc}
\usepackage{color}

\newcommand\edits[1]{ #1}

\newcommand{\der}{\mathrm{d}}

\begin{document}

\title[OMC length noise upper-limit]{\edits{Upper-limit on the Advanced Virgo
  output mode cleaner cavity length noise}}

\newcommand*{\LAPP}{Laboratoire d'Annecy-le-Vieux de Physique des
  Particules (LAPP), Université Savoie Mont Blanc, CNRS/IN2P3, F-74941
  Annecy, France
}

\author{R~Bonnand, M~Ducrot, R~Gouaty, F~Marion, A~Masserot, B~Mours, 
   E~Pacaud, L~Rolland, M~Wąs} 

\address{\LAPP} 
\ead{michal.was@lapp.in2p3.fr}

\date{\today}

\begin{abstract}
The Advanced Virgo detector uses two monolithic optical cavities at its
output port to suppress higher order modes and radio frequency
sidebands from the carrier light used for gravitational wave
detection. These two cavities in series form the output mode cleaner. 
We present a measured upper limit on the length noise of these
cavities that is consistent with the thermo-refractive noise
prediction of \SI{8e-16}{m/Hz^{1/2}} at \SI{15}{Hz}. The cavity length
is controlled using Peltier cells and piezo-electric actuators to
maintain resonance on the incoming light. A length lock precision 
of \SI{3.5e-13}{m} is achieved. These two results are combined to
demonstrate that the \edits{broadband} length noise of the output mode cleaner in the
\edits{10-\SI{60}{Hz} band} is at
least a factor 10 below other expected noise sources in the Advanced
Virgo detector design configuration.
\end{abstract}


\maketitle

\section{Introduction}
Advanced interferometric gravitational wave detectors, such as Advanced Virgo~\cite{aVirgo},
advanced LIGO~\cite{aLIGO}, or GEO-HF~\cite{GEOHFprogram} are making first 
detections or are about to start observations. 
All these detectors are using a special case of homodyne detection
called DC readout~\cite{Hild09, Fricke11} to extract the
differential arm length signal from the light at the interferometer
output, that is the carrier light. A crucial element in the
DC readout detection scheme is an output mode cleaner (OMC), \edits{which 
is a non-degenerate
optical cavity that transmits only the fundamental Gaussian mode at
the carrier frequency. The purpose is to keep only light which leaves
the interferometer due to a gravitational wave signal and remove higher} order modes caused by
interferometer mirror defects and radio-frequency sidebands used for
the control of auxiliary degrees of freedom. 

One drawback of this scheme is that any length noise of the OMC cavity is imprinted on the
transmitted light if the cavity length is not perfectly adjusted to
the carrier frequency. The coupling factor is proportional to the
root-mean-square of the difference between light carrier frequency and
cavity resonant frequency. 

A very low cavity length noise of a few \SI{e-17}{m/Hz^{1/2}} at
\SI{1}{Hz} has been obtained for rigid cavities~\cite{Numata04},
however these have no means to tune the cavity length to follow the
light frequency at the output of the interferometer. Although a scheme
has been proposed to remove the need for cavity
length actuator \cite{Vajente12}, all current detectors have actuators
on the OMC length that may introduce additional length noise. 

For advanced LIGO and GEO-HF the OMC is a 4-mirror bow-tie cavity with
one of the mirrors directly mounted on a piezo-electric actuator
(PZT)~\cite{aLIGO,Prijatelj11}, an upper limit of
\SI{2e-14}{m/Hz^{1/2}} on the cavity length noise introduced by this
PZT has been measured \edits{in the 1-\SI{7}{kHz} band}~\cite{Prijatelj11}. Advanced Virgo chose an
alternative, more compact design
with a single piece of fused silica forming a 4 surface bow-tie
cavity~\cite{aVirgo} \edits{(see figure~\ref{fig:setup})}, based on previous experience from Virgo~\cite{Beauville06}. The
cavity optical length can be controlled with two actuators, a Peltier cell
that thermally changes the refractive index in the cavity and hence
the optical length, and a PZT pressing the OMC transversely and
allowing a fast control of the refractive index but with a low dynamic
range.  This choice \edits{should have}
the advantage of reducing noise from mechanical vibration and from the PZT,
but has the drawback of light circulating in the substrate instead of
vacuum, which among other things introduces additional thermal noise~\cite{aVirgoTDR}. In
order to obtain sufficient light filtering without introducing 
high thermal noise and large scattered light losses, two
monolithic cavities are placed in series instead of a single cavity
with high finesse.

In this paper we present an upper limit on the Advanced Virgo OMC
length noise and the achieved precision of the OMC length control in a
table top measurement. By combining these two measurements we
derive the expected contribution of the OMC length noise
to the Advanced Virgo measurement noise.  In section~\ref{sec:OMC} we
discuss in detail the expected thermal length noise of the OMC and its
coupling to the gravitational wave measurement, then in
section~\ref{sec:setup} we describe the test measurement
setup. Section~\ref{sec:lengthNoise} presents the length noise upper
limit and section~\ref{sec:lockPrecision} the length control
precision.

\section{Advanced Virgo Output Mode Cleaner}
\label{sec:OMC}

The optimization of parameters for the Advanced Virgo OMC led to a
design composed of two monolithic fused silica cavities in series~\cite{aVirgoTDR},
which allows good radio frequency filtering with low finesse and
short cavities. Each OMC cavity is a single piece of fused silica with
an elongated hexagon shape. The cavity has an effective length $L =
\SI{0.124}{m}$ (half of the round-trip length) which corresponds to an
optical path length $l_0 = n L$ where $n$ is the fused silica
refractive index. The cavity finesse was measured to be $F \simeq
125$ \edits{and internal cavity losses at $\sim1.5$\%}~\cite{Ducrot14}. One of the cavity surfaces is curved, with a radius of
\SI{1.7}{m}, and the beam resonating in the cavity has a waist
$w_0 = \SI{321}{\micro\metre}$ located at the cavity input surface. 

The cavity couples the fluctuations in optical path length $l$
to power fluctuations $\delta P$ of the transmitted light 
\begin{equation}\label{eq:cavityTransExact}
  \frac{\delta P}{P_0} =  \frac{1}{1 + \left(\frac{2 F}{\pi}\right)^2 
    \sin^2\frac{2\pi l}{\lambda}
  } -1,
\end{equation}
where $\lambda=\SI{1064}{nm}$ is the laser wavelength. 
In practice the optical path fluctuations are dominated by low frequency
fluctuations, hence we decompose the cavity
length fluctuations into a large dynamic, low frequency (below
\SI{10}{Hz}) component with root-mean-square (RMS) $\Delta
l_\text{rms}$ and a small component $\delta l$ in the sensing
band of Advanced Virgo (\SI{10}{Hz}--\,\SI{10}{kHz})  
\begin{equation}
l = l_0 + \Delta l_\text{rms} + \delta l.
\end{equation}
Using this decomposition,
\eref{eq:cavityTransExact} can be approximated above
\SI{10}{Hz} by
\begin{equation}\label{eq:cavityTrans}
  \frac{\delta P}{P_0}  \simeq -32 \,F^2\, \frac{\Delta l_\text{rms}
      \delta l}{\lambda^2}.
\end{equation}

The corresponding noise on the gravitational wave signal is directly
given by the power fluctuations in
transmission of the OMC divided by interferometer response transfer
function $O_\text{TF}$. Hence the OMC length noise coupling into the gravitational
wave signal is
\begin{equation}\label{eq:hNoise}
  \delta h  = -32\, \sqrt{2}\, \frac{F^2 \Delta l_\text{rms}
      \delta l}{\lambda^2 O_\text{TF} },
\end{equation}
where the additional factor $\sqrt{2}$ comes from adding in quadrature
the length noise of the two OMC cavities, \edits{which is
expected to be caused by statistical fluctuations in
the substrate temperature and therefore independent.}

\begin{table}
  \centering
  \begin{tabular}{lcr}
    refractive index & $n$ & 1.44963 \\
    refractive index temperature dependence & $\beta$ &
                                                        \SI{-e-5}{K^{-1}}\\
    density & $D$ & \SI{2200}{kg/m^3}\\
    thermal conductivity & $\kappa$ & \SI{1.38}{W m^{-1} K^{-1}} \\
    temperature & $T$ &  \SI{300}{K} \\
    specific heat & $C$ & \SI{746}{J K^{-1} kg^{-1}}
                          
  \end{tabular}
  \caption{Fused silica parameters for thermo-refractive noise
    calculation \eref{eq:thermoRefractive}}
  \label{tab:params}
\end{table}

\edits{Indeed,} compared to a 4 mirror cavity design, a monolithic cavity has
additional thermal fluctuations from the medium in which light
circulates. Given that the thermal expansion coefficient of fused silica
$\alpha = \SI{5e-7}{K^{-1}}$ is small compared to the changes of the
refractive index as a function of temperature
$\frac{\der n}{\der T} = \beta = \SI{-e-5}{K^{-1}}$, the dominant
thermal length noise is thermo-refractive noise. There is also a
contribution from Brownian noise and from all coatings thermal noises, but
these are also negligible compared to thermo-refractive noise.

Thermo-refractive noise has been 
computed as a function of frequency $f$ for an infinite plane of
thickness $L$ in
appendix E of \cite{Braginsky04} and can be rewritten as
\begin{equation}\label{eq:thermoRefractive}\fl
  \delta l_\text{thermo-refractive}(f) \simeq \frac{2 \beta T \sqrt{L k_\text{B}
    \kappa}}{D C \sqrt{\pi}} \frac{1}{ (w_0/\sqrt{2})^2 2\pi f }
\left[\int_0^\infty \frac{x \der x }{1 + \frac{4 r_\text{th}^2}{w_0^2} x^2} e^{-x}\right]^{1/2},
\end{equation}
where $k_\text{B}$ is the Boltzmann constant, $r_\text{th} =
\sqrt{\kappa/(2 \pi f D C)}$ is the thermal path length, \edits{$L =
\SI{0.124}{m}$ is half of the cavity round-trip length} and the values of the different parameters for fused silica are
given in table~\ref{tab:params}. The integral term equates to 1 in the
adiabatic limit where $r_\text{th} \ll w_0$. At low frequencies where
$r_\text{th} \gg w_0$, the integral term is $\propto f$, hence the
thermo-refractive noise remains bounded at low frequency. \edits{The
transition between the two regimes occurs at $f\sim\SI{1.3}{Hz}$ for
the OMC cavity.} The same result has been obtained
more recently for a finite cavity of cylindrical geometry \cite{Heinert11},
which restrains its validity to the range where $r_\text{th}$  is much larger than
the wavelength and much smaller than the transverse size of the cavity
(\SI{1}{cm}). This corresponds to a valid frequency range of
\SI{10}{mHz}--\,\SI{100}{kHz}, which covers well the frequency range of
interest here. The principle of these computations has been confirmed
by measuring thermo-refractive noise in a very different geometry of
whispering-gallery mode of microspheres~\cite{Gorodetsky04}.

In section \ref{sec:lengthNoise} we measure that the cavity length noise
at \SI{10}{Hz} is not larger than the one predicted by \eref{eq:thermoRefractive}.

\section{Experimental setup}
\label{sec:setup}

\begin{figure}
  \includegraphics[width=\textwidth]{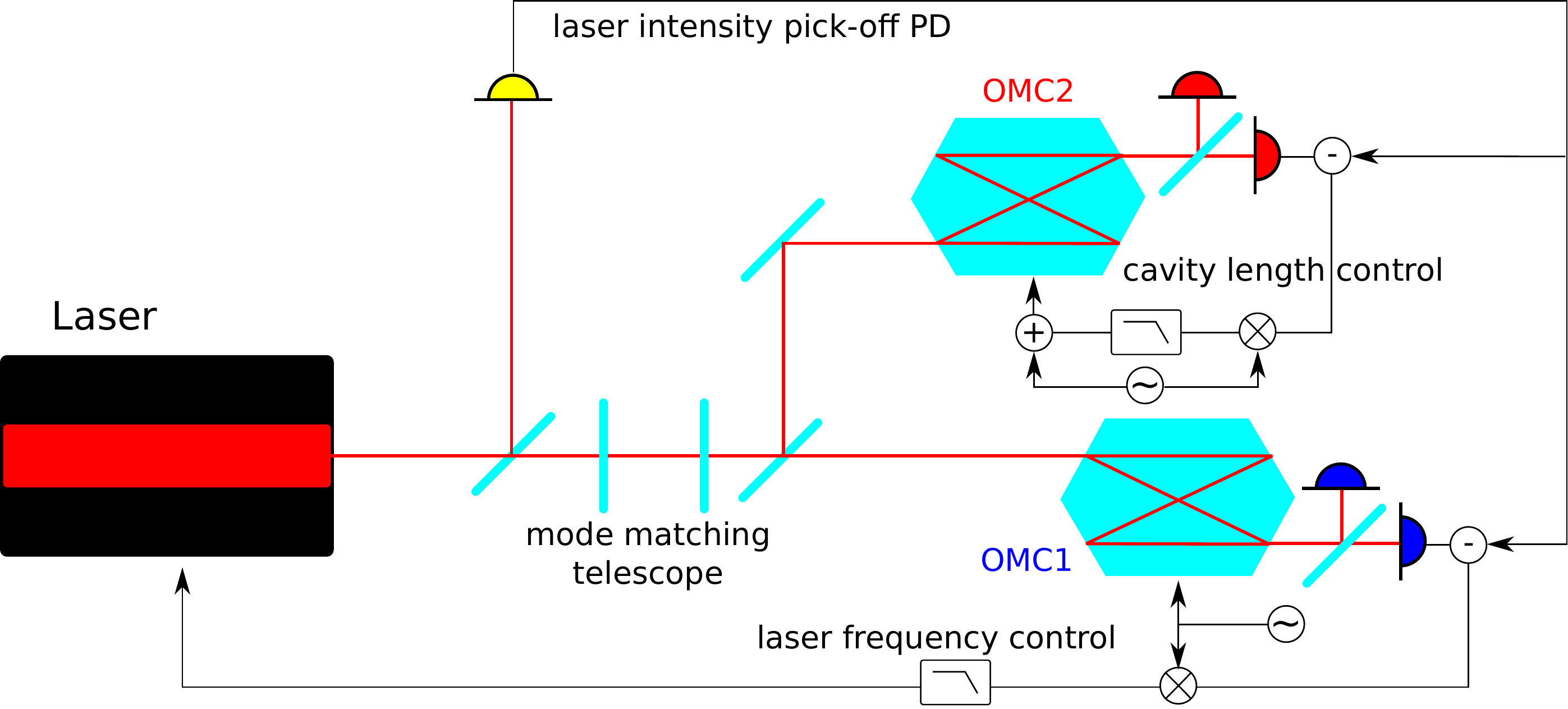}  
  \caption{Schematic of the optical test setup (not to scale). The two
    OMC cavities OMC1 and OMC2 are set in parallel and each receives
    45\% of the laser beam, which in transmission of each cavity is
    split equally onto an in-loop and an out-of-loop PD. The flow of
    signal for the subtraction of laser intensity noise, the laser
    frequency control and the cavity length control is shown by
    arrows.}
  \label{fig:setup}
\end{figure}

The OMC cavity length fluctuations are measured using a dedicated
optical test setup. The setup is located on a passively isolated optical bench,
enclosed in aluminium and plexiglas covers placed on a tubular
structure to prevent beam jitter from the clean room air flow. A
schematic of the optical layout is shown on
figure~\ref{fig:setup}. \edits{A small fraction (10\%) of the light} from a \SI{2}{W} Mephisto
laser~\cite{MephistoLaser} at
\SI{1064}{nm} is picked-off to a photo-diode (PD) to measure the laser
intensity noise; the main part of the beam is 
matched with a telescope to two cavities set in parallel. This is
different from the Advanced Virgo case where the two cavities are
placed in series, and allows a simple measurement where light seen by
one cavity is not directly affected by the other. For each
cavity two PDs measure the transmitted power, which is between
\SI{30}{mW} and \SI{80}{mW} depending on the PD. 

To obtain an error signal for the cavity length, a dithering sine-wave,
at a dozen kHz is applied to each cavity by a PZT. The cavity
length error point is the PD signal demodulated at the dither
frequency. This signal is limited by the laser intensity noise at the
dither frequency. With the active power stabilization loop enabled (``noise eater''~\cite{MephistoLaser}), the
laser intensity noise at a dozen kHz has a relative intensity
$\sim$\SI{2e-7}{Hz^{-1/2}}.
This laser intensity is measured by a pick-off
PD and subtracted before demodulation.

The length error point is calibrated by scanning linearly the laser
frequency over several cavity free spectral ranges, each free spectral
range corresponding to a cavity length change of $\lambda/2$, as can be
seen from \eref{eq:cavityTransExact}. The measurement
has an absolute statistical error of 2\,--3\,\% and the systematic
error from using the cavity length as a reference is less than 1\%. Consequently the relative
calibration is adjusted by several percent to obtain a good
cancellation of common frequency noise between the two cavities. This
cancellation is described in section~\ref{sec:lengthNoise}.

\section{Cavity length noise measurement}
\label{sec:lengthNoise}

In our measurement the laser frequency noise is dominant. Hence we
lock the laser onto the length of OMC1 with a bandwidth of
$\sim\SI{250}{Hz}$ to reduce the frequency noise, and leave the OMC1
cavity length free. The OMC2 length is
locked onto the laser frequency using Peltier cells and the PZT
actuator with a loop unity gain frequency around $\SI{0.2}{Hz}$.

\begin{figure}
  \centering
  \includegraphics[width=0.9\textwidth]{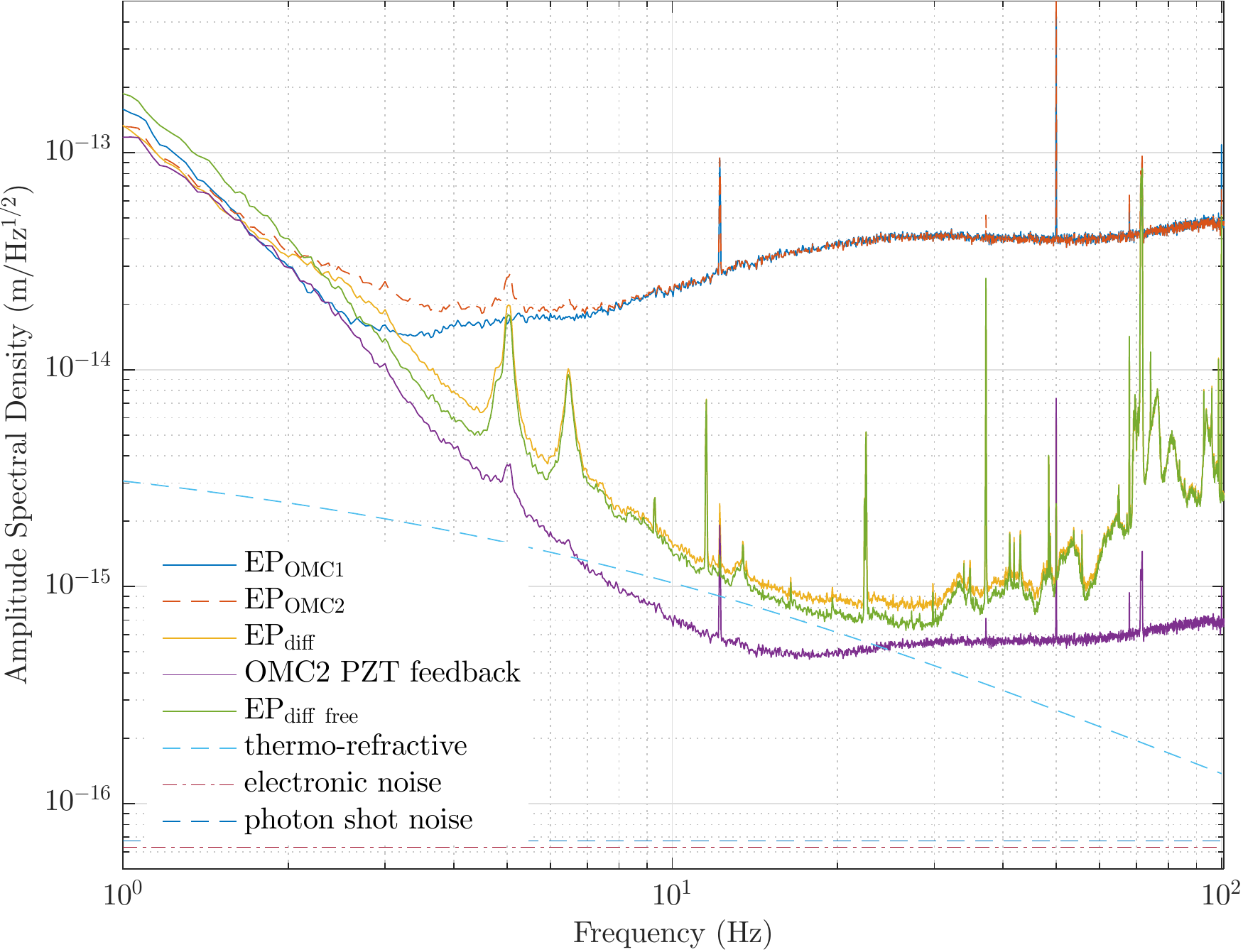}
  \caption{Differential OMC length noise measurement \edits{over 1.5 hours of
    data}. Shown are the
    calibrated length error points for OMC1 and OMC2, their difference
  \eref{eq:EP_diff}, and the difference with the OMC2
  PZT feedback subtracted \eref{eq:EP_diff_free}. For
  reference the expected thermo-refractive noise is shown along with
  electronic and photon shot noise. }
  \label{fig:lengthNoise}
\end{figure}

Laser frequency noise $\delta \nu$
couples to transmitted light fluctuation $\delta P$ in the same way as
cavity length noise as shown in \eref{eq:cavityTransExact}.
The spectra of the two cavities length error points are shown on
figure~\ref{fig:lengthNoise} and are dominated by laser frequency
noise. This frequency noise is equivalent to cavity length noise
$\delta l_\text{laser}$ using the relation
\begin{equation}\label{eq:freqLength}
  \frac{\delta \nu}{\nu} = -\frac{\delta \lambda}{\lambda} = \frac{\delta l_\text{laser} }{l_0},
\end{equation}
with $\nu=\SI{2.82e14}{Hz}$ the light frequency. 
With this notation and dismissing other noise contributions the cavity calibrated
error point can be written as
\begin{equation}
  \text{EP}_\text{OMC} = \delta l_\text{OMC} + \delta l_\text{laser}.
\end{equation}
Hence the difference between the two cavities 
\begin{equation}\label{eq:EP_diff}
  \text{EP}_\text{diff} = \text{EP}_\text{OMC1} - \text{EP}_\text{OMC2} = \delta l_\text{OMC1}  - \delta l_\text{OMC2}
\end{equation}
should be a good measure of the differential length, free of the
common laser frequency noise. This is a true representation of the
cavity differential length above \SI{10}{Hz} as the OMC1 length is
free and the OMC2 length control loop has a gain well below 1.  However, some of the
laser frequency noise is re-injected by PZT feedback of the OMC2
length control, as the loop gain at 10-$\SI{100}{Hz}$ is $\sim0.02$
and not zero. This means that in addition to OMC2 free cavity length
noise there is a loop feedback contribution
$\delta l_\text{OMC2} = \delta l_\text{OMC2, free} + \delta
l_\text{OMC2, PZT}$.  The latter can be easily subtracted as the
feedback voltage is known and the cavity has a flat response of
$\chi=\SI{2.6e-11}{m/V}$ at these frequencies, hence we define the loop
corrected differential length as
\begin{equation}\label{eq:EP_diff_free}\fl
  \text{EP}_\text{diff free} = \text{EP}_\text{OMC1} -
  (\text{EP}_\text{OMC2} - \chi \text{PZT}_\text{OMC2}) 
  = \delta l_\text{OMC1,free}  - \delta l_\text{OMC2,free}.
\end{equation}

These noise curves are shown on figure~\ref{fig:lengthNoise} \edits{averaged
over 1.5 hours of data, and we have checked that this noise level is
  stationary at the few minutes time scale. Also shown
are the PD electronic noise measured with the laser
switched off and the expected photon shot-noise.} For comparison the
thermo-refractive noise of the two cavities from \ref{eq:thermoRefractive} added in quadrature is
also shown.
The measured cavity
differential length shows many lines due to mechanical resonances \edits{of
components on the optical bench, especially above \SI{60}{Hz} the measurement is
completely spoiled.} At lower frequencies,
the broad lines at $\SI{5}{Hz}$ and $\SI{6.5}{Hz}$ come from the tubular
posts that hold plexiglas covers to prevent air-flow, and their
coupling depends on the torque applied to the PZT clamping.  
Below
$\SI{10}{Hz}$ there are large fluctuations that are not understood.
Nonetheless in the 10-$\SI{20}{Hz}$ band the measured noise is within 10\%
of the thermo-refractive prediction, and is an upper-limit that
no length noise is larger than this prediction.

\section{Length control precision}
\label{sec:lockPrecision}

\begin{figure}
  \centering
  \includegraphics[width=0.9\textwidth]{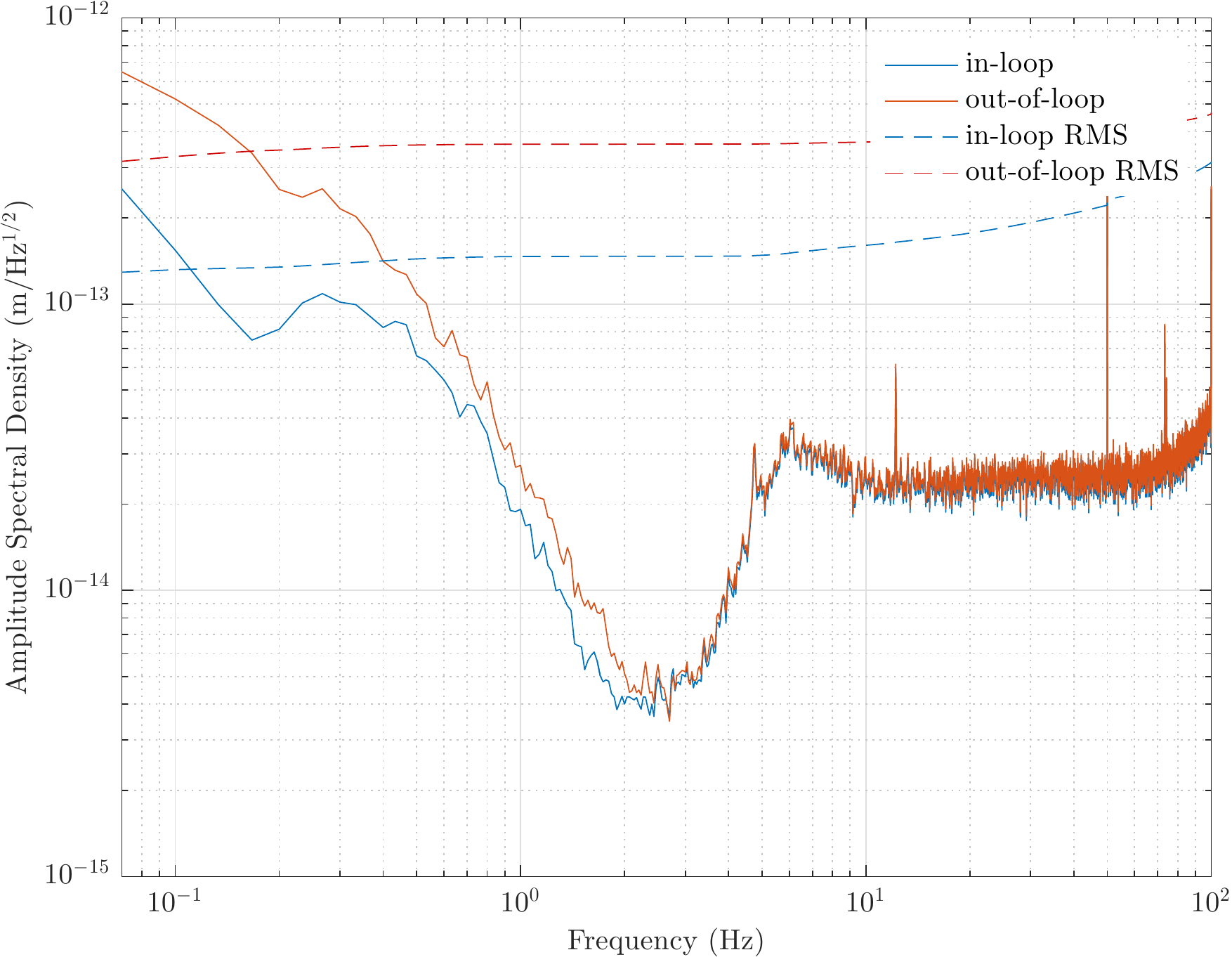}
  \caption{Spectra of the in-loop and out-of-loop measures of the OMC2
  cavity length. The integrated low-frequency RMS of both signals are
  also shown. Above \SI{5}{Hz} the signal is dominated by laser
  frequency noise, which is slightly lower than in
  figure~\ref{fig:lengthNoise} due to higher gain laser frequency
  control. 
  Below that frequency the out-of-loop length RMS is \SI{3.5e-13}{m}. }
  \label{fig:lockPrecision}
\end{figure}

The cavity length fluctuations couple to the gravitational wave
detector sensitivity through \edits{\eref{eq:hNoise}, where the lock
precision $\Delta l_\text{rms}$ is a critical parameter}. In
section~\ref{sec:lengthNoise} we used a low gain on the OMC2 cavity
length control to limit re-injecting frequency noise with the loop and
to simplify the result interpretation. To improve the lock precision, the
OMC is operated with a factor 10 higher loop gain and a unity
gain at a few Hz. As in the previous section the loop feedback is dominated by the
Peltier cells temperature feedback below \SI{0.1}{Hz} and by the PZT
feedback above this cross-over frequency. Figure~\ref{fig:lockPrecision} shows the OMC2
error point spectrum for a high gain loop using the in-loop and
out-of-loop PD. This achieves  $\Delta l_\text{rms} = \SI{3.5e-13}{m}$ for the
out-of-loop signal, over 3 times lower than the design requirement~\cite{aVirgo}.

In Advanced Virgo the precision of the laser lock onto the reference
cavity is $\sim\SI{1}{Hz}$~RMS~\cite{aVirgo}, which through
\ref{eq:freqLength} is equivalent to $\SI{4.4e-16}{m}$, that
is a laser frequency control orders of magnitude better that the one
achieved here. Therefore in the real case the laser frequency should
not be dominant as in the test setup presented here, and the OMC lock
precision should be easily reproduced even with a lower control loop
gain.

\begin{figure}
  \centering
  \includegraphics[width=0.9\textwidth]{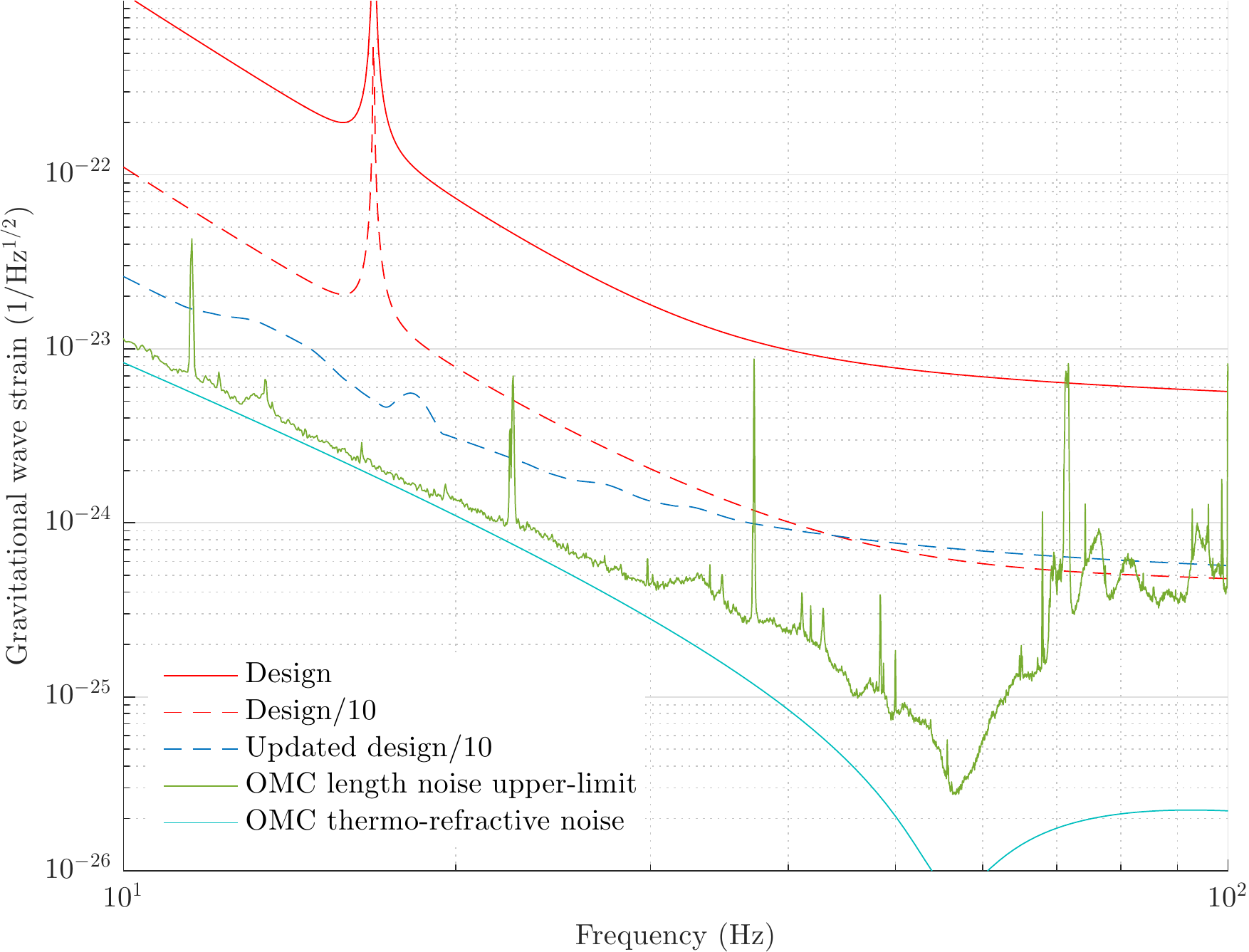}
  \caption{Projection of the upper-limit on OMC length noise from
    figure~\ref{fig:lengthNoise} onto the Advanced Virgo design noise
    curve assuming a lock precision of \SI{3.5e-13}{m} and a detuned
    signal recycling configuration. The original design and updated
    design Advanced Virgo noise curves are shown.  }
  \label{fig:noiseProjection}
\end{figure}

Using \ref{eq:hNoise} we combine the achieved lock precision
$\Delta l_\text{rms}$ and the measured upper limit on cavity noise
$\delta l$ shown in figure~\ref{fig:lengthNoise} to obtain the
expected contribution of OMC length noise to the Advanced Virgo
measurement noise shown on figure~\ref{fig:noiseProjection}. Below
\SI{60}{Hz} this expected \edits{broadband} noise is at least 10 times smaller than the
Advanced Virgo design sensitivity~\cite{aVirgoTDR} or the updated
sensitivity expectation using more accurate suspension thermal noise
models~\cite{aVirgo}. Above \SI{60}{Hz} the measured OMC length noise
upper limit is dominated by mechanical resonance of the measurement setup,
nonetheless it remains below the design sensitivity curve. \edits{In Advanced
Virgo these resonances should not be present as the OMC will be
placed on a suspended bench in vacuum. }

\edits{The prominent lines at \SI{11.53}{Hz}, \SI{22.53}{Hz} and
  \SI{37.23}{Hz} are narrow, with respective linewidths of
  \SI{30}{mHz}, \SI{100}{mHz} and \SI{30}{mHz}. Hence regardless of
  whether these are real length noise or a sensing noise of this
  particular measurement,
  these lines will not have an impact on the broadband gravitational
  wave sensitivity that is relevant for most gravitational wave
  signals, such as coming from compact binary coalescences~\cite{O1BBH}.}

Note that in the initial broadband configurations~\cite{aVirgoTDR} the
constraints on the OMC length noise are $\sim10$ weaker than in the
configuration optimized for binary neutron star detection.  Indeed,
without signal recycling or with signal recycling tuned the optical
gain at \SI{10}{Hz} is about an order of magnitude higher, hence the
OMC length noise coupling in these cases is 10 times smaller at low
frequency than presented here.

\section{Conclusions}
\label{sec:conclusion}
We have measured an upper limit on the Advanced Virgo OMC cavity
length noise. In the 10-\SI{20}{Hz} band this upper limit is
consistent with the thermo-refractive noise prediction, confirming
that there is no other significant length noise in this cavity design.

We achieved a lock precision of
$\Delta l_\text{rms} = \SI{3.5e-13}{m}$, which is a factor 17 better
than obtained previously for the Virgo OMC~\cite{aVirgoTDR}.
Combining these two results we have shown that the OMC length noise
contribution to the Advanced Virgo measurement should be at least a
factor 10 below other expected noise sources in the 10-\SI{60}{Hz}
band, which is most challenging for technical noises. 
This gives confidence
that the OMC noise should not be a limiting factor in the forthcoming
Advanced Virgo observations. 

The achieved gap between the expected OMC noise contribution and
Advanced Virgo noise opens the prospects of further parameter
optimization depending on the issues encountered during the first
Advanced Virgo operations. For example the radio-frequency sideband
filtering could be improved by an order of magnitude by increasing the
OMC finesse by a factor 2. This would increase the thermo-refractive
noise by a factor 4 but it would still remain a factor 10 below other
expected noises.

\ack

We would like to thank our Advanced Virgo collaborators for numerous
discussions that lead to the design and optimization of the OMC
cavities discussed here.

\section*{References}

\bibliographystyle{unsrt}

\end{document}